# Music Similarity Calculation of Individual Instrumental Sounds Using Metric Learning

Yuka Hashizume\*, Li Li† and Tomoki Toda‡

\* Nagoya University, Nagoya, Japan
E-mail: hashizume.yuuka@g.sp.m.is.nagoya-u.ac.jp

† Nagoya University, Nagoya, Japan
E-mail: li.li@g.sp.m.is.nagoya-u.ac.jp

‡ Nagoya University, Nagoya, Japan
E-mail: tomoki@icts.nagoya-u.ac.jp

*Abstract*—The criteria for measuring music similarity are important for developing a flexible music recommendation system. Some data-driven methods have been proposed to calculate music similarity from only music signals, such as metric learning based on a triplet loss using tag information on each musical piece. However, the resulting music similarity metric usually captures the entire piece of music, i.e., the mixing of various instrumental sounds sources, limiting the capability of the music recommendation system, e.g., it is difficult to search for a musical piece containing similar drum sounds. Towards the development of a more flexible music recommendation system, we propose a music similarity calculation method that focuses on individual instrumental sound sources in a musical piece. By fully exploiting the potential of data-driven methods for our proposed method, we employ weakly supervised metric learning to individual instrumental sound source signals without using any tag information, where positive and negative samples in a triplet loss are defined by whether or not they are from the same musical piece. Furthermore, assuming that each instrumental sound source is not always available in practice, we also investigate the effects of using instrumental sound source separation to obtain each source in the proposed method. Experimental results have shown that (1) unique similarity metrics can be learned for individual instrumental sound sources, (2) similarity metrics learned using some instrumental sound sources are possible to lead to more accurate results than that learned using the entire musical piece, (3) the performance degraded when learning with the separated instrumental sounds, and (4) similarity metrics learned by the proposed method well produced results that correspond to perception by human senses.

## I. INTRODUCTION

The amount of music available on the Internet is enormous and continues to grow. Under such circumstances, it is impossible to listen to all the music in the world to find users' favorite music. Therefore, a music information retrieval (MIR) technique, such as a music recommendation system, is necessary to help users find their favorite music efficiently, and the development of a suitable criterion for measuring music similarity is essential.

One of the typical methods for calculating the similarity between musical pieces is to utilize the user's listening history [1]. A collaborative filtering technique is one of the most successful approaches. This method assumes that users who have rated some items similarly or behaved in the same way will also rate other items similarly. Therefore, scores for unseen music can be predicted from the scores rated by other users with similar behavior. However, one limitation is that newly released music may be rarely recommended until a certain amount of listening history has been recorded. Another problem is that users rarely listen to music that is not well known since popular music can generally get more ratings.

Different from the collaborative filtering, another main branch of conventional methods is content-based approaches, which recommend music on the basis of content similarity instead of user's behavior. The content-based similarity is generally obtained by extracting feature representations from music signals and calculating the similarity or distance between the representations. Before the advent of deep learning, low-level features such as chord progression or tempo and high-level features such as manually designed acoustic features were used as the feature representations [2]. The similarity between them is then measured by using a distance criterion, such as cosine distance and Euclidean distance. Content-based methods avoid the problem associated with collaborative filtering since they do not require human ratings. However, their performance depends on handcrafted features and the selected distance criterion, which do not always work well owing to the lack of generalization capability.

With the advent of deep learning, data-driven feature extraction has shown to be effective in improving the performance of MIR systems [3], [4]. For example, [4], [5] have proposed to extract the feature representations from the middle layer of a genre classifier. Some methods have been proposed to learn the feature representations by metric learning using tag information such as tags given by humans [6], artist tags [7], genre tags [8], and tags using zero-shot learning [9]. These conventional methods measure the music similarity by taking the entire musical piece, i.e., the mixing of various instrumental sound, into consideration. However, since the perspective of a musical piece the users want to listen to varies from user to user, measuring a musical piece only from a single perspective is insufficient. To achieve more flexible MIR systems, calculating music similarities using more varieties of feature representations of musical pieces is necessary.

In this paper, to achieve more flexible calculation of music similarity, we propose a music similarity calculation method that focuses on individual instrumental sound sources in a musical piece. By fully exploiting the potential of data-driven feature extraction techniques for our proposed method we extract the feature representations by using weakly supervised metric learning with track information instead of using any tag information. In the proposed method, the similarity is calculated using each instrumental sound source instead of the entire musical piece. Furthermore, assuming that individual instrumental sources are difficult to obtain in practice, we apply instrumental source separation to obtain each source from the original mixed musical piece and evaluate the proposed method with separated signals. By conducting the objective and subjective experiments, we mainly investigate the following three questions: (1) whether unique and useful similarity metrics can be learned for different instruments, (2) whether separated instrumental source signals can be used for similarity calculation in the proposed method, and (3) whether the learned similarity metrics can produce results that correspond to the perception of human senses.

## II. RELATED RESEARCH

### A. Feature extraction for computing similarity between musical pieces

Li et al. [11] proposed a method of extracting the feature representations using a convolutional neural network (CNN), which took mel frequency cepstral coefficients (MFCCs) as a network input and output the predicted genre. Latent features used as an input of the last layer were extracted as the feature representations from MFCCs. Recently, there has been an increasing number of studies using mel-spectrograms as the input instead of MFCCs. For example, Fathollahi and Razzazi [5] split a musical piece into three-second segments and converted each of them into a corresponding mel-spectrogram, which was then used as the input for CNN. In [5], they compared performances achieved by varying a training setting, such as segment lengths of 3, 5, and 10 s, and those having an overlap or not. The experimental results showed that the use of the segments of 3 s with overlap produced the best results.

### B. Metric learning of similarity between musical pieces using triplet loss

In deep metric learning [12], it is aimed to automatically construct a distance metric for a specific task in a machine learning manner, which generally can find distance metrics more suitable for the task than handcrafted ones. With a triplet loss [10], a distance metric is trained with a triplet of samples, where one is considered as an anchor and the other two are considered as positive and negative samples. Here, the positive sample should be more similar to the anchor than the negative one. Lee et al. [8] proposed computation of the similarity between musical pieces by metric learning using the triplet loss. They proposed the track-based similarity which is learned by triplets of samples extracted using only track information. Namely, segments from the same track as the anchor's one are defined as positive samples, and those from different tracks from the anchor's one are defined as negative samples.

If we use $x_i^{(a)}$, $x_i^{(p)}$, and $x_i^{(n)}$ to denote the $i$th anchor, positive sample, and negative sample, respectively, the triplet $t_i$ is constructed as a set of $\{x_i^{(a)}, x_i^{(p)}, x_i^{(n)}\}$, where $i = 1, \ldots, I$ denotes the index of training samples. The triplet loss is defined as

$$\mathcal{L}(t_i) = \max\{d(x_i^{(a)}, x_i^{(p)}) - d(x_i^{(a)}, x_i^{(n)}) + \Delta, 0\}, \quad (1)$$

where $d$ is a distance function for measuring the distance between two audio samples, such as the Euclidean distance or cosine distance, and $\Delta$ is a margin value, which defines the minimum distance between the positive and negative samples.

## III. PROPOSED METHOD

To achieve a highly flexible MIR system by making it possible to handle music similarity from more various perspectives than the conventional method focusing on only the entire musical piece, we propose a music similarity calculation method that focuses on each instrumental sound source as a partial element of a piece of music. An overview of the proposed method is illustrated in Fig.1.

In the proposed method, we apply metric learning with the triplet loss to individual instrumental sound sources, e.g, drum, piano, and guitar. Different from genre, artist, and mood tags, the annotation of tags that represent music similarity focusing on each instrumental sound is very human-resource-intensive. Therefore, we use the track-based similarity described in Section 2.2 to define the positive and negative samples. Namely, we define a segment from the same musical piece as that of the anchor sample as a positive sample and a segment from a different musical piece as a negative sample.

A network consisting of convolutional layers and a fully connected layer is used to extract the feature representation for the music similarity calculation in metric learning. The network architecture is shown in Fig. 2. Different networks are separately trained for individual instrumental sounds. We use cosine distance between the feature representations to measure the distance between the anchor and positive samples and that between the anchor and negative samples, namely, $d(x_i^a, x_i^p)$ and $d(x_i^a, x_i^n)$, respectively. In addition to the use of recorded instrumental sound sources, we also apply the proposed method to separated instrument signals, which are extracted from the mixed music signals by using an instrumental source separation method assuming that the recorded instrumental sound tracks are not always available in practice.

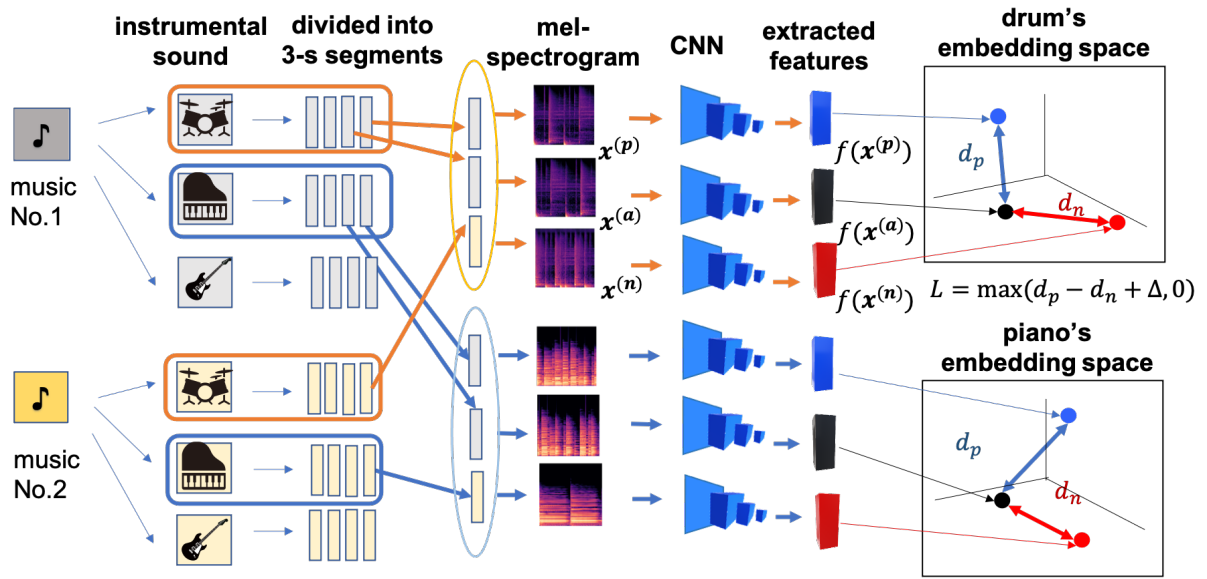

Fig. 1. *Overview of the proposed method. Feature representations for individual instrumental sounds are extracted separately using metric learning with triplet loss. The $x^{(a)}$, $x^{(p)}$, and $x^{(n)}$ denote the anchor, positive, and negative sample, respectively.*

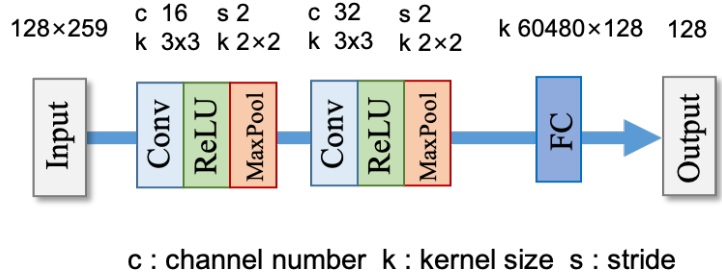

Fig. 2. *Network architecture of CNN. The 'c', 'k', and 's' denote the channel number, kernel size, and stride, respectively. "Conv" and "FC" denote convolutional layer and fully connected layer, respectively. The numbers above input and output are their data sizes.*

## IV. EXPERIMENTAL EVALUATION

### A. Experimental conditions

The dataset we used is slakh [13], which contains not only sound sources of a mixture of various instrumental sounds (hereafter referred to as the mixed sound) but also sound sources of each instrumental sound (hereafter referred to as the original instrumental sound). Note that the musical pieces in this dataset do not contain vocals. We used the original instrumental sounds of drums, bass, piano, and guitar in the proposed method to measure music similarity focusing on the individual instrumental sounds. In addition, we extracted drums, bass, and piano sounds from the mixed sound using an instrumental sound source separation method, "spleeter" [14] in the Python library that could separate a music signal into drums, bass, piano, vocals, and others, and also used those three separated sounds (hereafter referred to as the separated instrumental sounds) in the proposed method to investigate the effects of using the separated instrumental sounds on music similarity calculation. Spleeter is an instrumental sound source separation method using 12-layer U-nets [15]. Its separation performance for the separated sounds used in this experiment is shown in Table I. As a reference, we also calculated music similarity using the mixed sound. This corresponds to the conventional method. The sampling rate for all data was 44.1 kHz.

We used 180 musical pieces from the dataset to extract feature representations in metric learning, split each musical piece into three-second segments with 50% overlap, and used the first 40 segments in each musical piece, excluding the silent parts. In total, 7200 training segments were extracted. For testing, 19 musical pieces were used, divided into segments in the same way, and all segments in each musical piece were used except for the silent parts. These 199 musical pieces in total were selected from the data set as having no less than 40 segments in all instrumental sounds when divided by the method described above, eliminating duplication. The extracted segments were converted into mel-spectrograms

TABLE I

*Separation performance for the separated sounds used as training data and test data in the experiment. Note that the SDRs before the separation of train data are -58.8, -58.8, and -57.3, respectively, when drums, bass, and piano are used as references. Similarly, the SDRs before the separation of test data are -63.2, -60.0, and -64.2. These values were calculated using "fast_bss_eval" [16].*

| instrument | SDR | SIR | SAR |
|---|---|---|---|
| | train / test | train / test | train / test |
| drums | -13.9 , -13.7 | 20.7 , 21.4 | -13.8 , -13.7 |
| bass | -16.24 , -15.5 | 8.7 , 9.7 | -14.9 , -14.7 |
| piano | -15.2 , -14.7 | 7.8 , 8.4 | -13.9 , -13.7 |

used as the input. Arbitrary anchors were selected from the mel-spectrograms, and metric learning based on the triplet loss was performed. The CNN in Fig. 2 was trained to extract a 128-dimensional embedding vector from a mel-spectrogram as a feature representation for measuring music similarity. The embeddings will be normalized to Euclidean norm of 1. The margin of the loss function was set to 0.2. A batch size was set to 64. The number of epochs was set to 150. We trained the CNN five times by changing initial settings, conducted evaluations described below using each of them, and averaged results over these five trials.

### B. Evaluation method

As experimental evaluations, we assessed whether the feature representations were well learned, whether the similarity based on each instrumental sound was a unique metric, and whether the similarity corresponded to perception by human senses.

*1) Objective evaluation:* In general, feature representations appropriate for a recommendation system need to satisfy two properties: (1) feature representations of similar items, i.e., segments from the same musical piece, are close to each other, (2) those of dissimilar items, i.e., segments from different musical pieces, are far apart from each other according to the degree of dissimilarity. To evaluate the learned feature representations, we used the accuracy of music IDs inferred using close feature representations. Specifically, we used the K-nearest neighbor (kNN) method to infer the music IDs of the test segments. The music IDs of all test segments except the one to be inferred were assumed to be known. We embedded all test segments into the learned feature representation space and predicted the music ID of each test segment by a majority vote using the IDs of the top five nearest test segments. The averaged accuracy rate of the entire test dataset over five trials was calculated for each feature representation space learned by using the mixed sound, the original instrumental sounds, or the separated instrumental sounds.

The aim of the proposed method is to construct unique similarity metrics when focusing on the individual instrumental sounds. To evaluate this point, we visualized distance matrices over centroid feature representations of the 19 test musical pieces, where the centroid feature representations were obtained by averaging feature representations over all segments from the same musical piece. We compared the averaged distance matrices calculated with the music similarity metrics learned using the mixed sound and the original instrumental sounds to investigate how different they were. In addition to visualization, we quantified the difference in music similarity metrics using a correlation coefficient and Spearman's rank correlation coefficient [17]. For a correlation coefficient, we vectorized the elements in the upper triangular part excluding the diagonal part of the averaged distance matrix for each of the original instrumental sounds and the mixed sound, and we calculated a correlation coefficient between two vectors of each pair of the averaged distance matrices. For Spearman's rank correlation coefficient, we ranked the musical pieces for each test musical piece using the averaged distance values, i.e., each column of the averaged distance matrix, and calculated Spearman's rank correlation coefficient between two ranks of each pair of the averaged distance matrices. The coefficients were calculated for all test musical pieces and were averaged over them.

*2) Subjective Evaluation:* To evaluate whether or not the learned similarity metrics focusing on individual instrumental sound sources (i.e., the proposed method) or the entire piece of music (i.e., the conventional method) can find perceptually similar segments in terms of the focused perspective than those learned by other types of sound, we conducted a subjective evaluation. Thirteen listeners participating in the experiment were asked to listen to audio sets that included three audio clips, consisting of an anchor and two candidates, and to select which one was more similar to the anchor. We also asked the listeners whether they were confident in their choice. In the listening experiment, we used the original individual sound source signals rather than the mixed music signal as the audio clips in order to help listeners easily focus on each sound source. We provided 40 audio sets to each listener, which were collected in the following manner.

We used the similarity between the centroid of the feature representations of each musical piece as the music similarity, and the top similar musical pieces for each musical piece were used in the evaluation. A 10-second sample was used for the evaluation, since it was not possible in time to have the listeners listen to a whole song. To evaluate each of the music similarity metrics learned by the original instrumental sounds, including drums, bass, piano, and guitar, and the mixed sounds, we made eight valid audio sets for each case (total 40 sets). For clarity of explanation, let us first consider the case of drums. For each audio set for evaluating the music similarity metric focusing on drums, we randomly selected a musical piece to be the anchor. A positive musical piece was then randomly selected from the first and second similar musical pieces to the anchor found in the drum sounds similarity. A negative musical piece was randomly selected from the first and second similar musical pieces to the anchor found in the mix sounds similarity. However, the audio set was considered invalid if at least one of the two candidates for positive and negative musical piece overlapped. Note that drums sounds of the musical pieces were presented to the listeners as the valid audio sets in the case of drums. In the case of the mixed sound, the positive musical piece was selected in the mixed sound similarity, and the negative musical piece was selected in a random instrumental sound similarity, and the mixed music sounds of the valid audio set were presented to the listeners. Therefore, the effectiveness of the learned similarity metrics can be confirmed when the candidate corresponding to the positive sample in the audio set is selected.

TABLE II
*kNN-based classification accuracy using each feature representation. The "original" and "separated" columns indicate the original and separated instrumental sounds, respectively. "mix" denotes the mixed sounds. The column "variance" shows the variance over the five trials.*

| original | accuracy[%] | variance | separated | accuracy[%] | variance |
|---|---|---|---|---|---|
| mix | 90.52 | 6.1e-4 | | | |
| drums | 93.68 | 4.5e-5 | drums | 76.92 | 1.4e-3 |
| bass | 84.63 | 3.9e-4 | bass | 41.41 | 1.4e-3 |
| piano | 92.17 | 3.4e-4 | piano | 65.69 | 1.4e-3 |
| guitar | 87.59 | 5.4e-4 | | | |

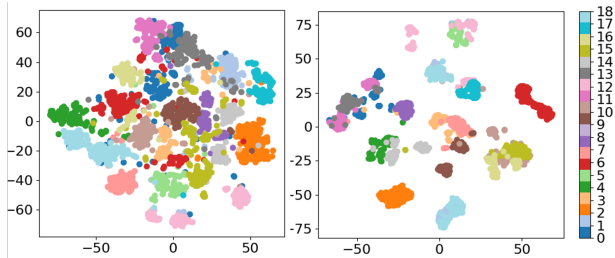

Fig. 3. *An example of visualized feature representations focusing on the mixed sound (left) and the drums sound (right). The numbers on the right side of the color bars show music IDs of 19 test musical pieces. Segments from the same musical piece are plotted with the same color in both diagrams.*

### C. Result

*1) Objective evaluation:* The accuracy rate of the predicted music IDs is shown in Table II. We found that high accuracies of about 90% to 95% were obtained in the cases of the mixed and original instrumental sounds. Accuracies obtained with the original drums and piano sounds were higher than those obtained with the mixed sounds, which indicated that similarity metrics learned using some instrumental source sounds could lead to more accurate results than those learned using the entire piece of music. On the other hand, accuracies of less than 80% were obtained in the cases of the separated instrumental sounds. One possible reason causing this degradation is the lower audio quality of the separated sounds than of the original ones caused by artifacts and residual components from other instruments. It is worthwhile to note that significantly low values of variance were achieved in all cases, indicating that the similarity metrics could be stably learned with different initial network parameters using the triplet loss. An example of the feature representation spaces is shown in Fig. 3. We used t-SNE [18] to compress the

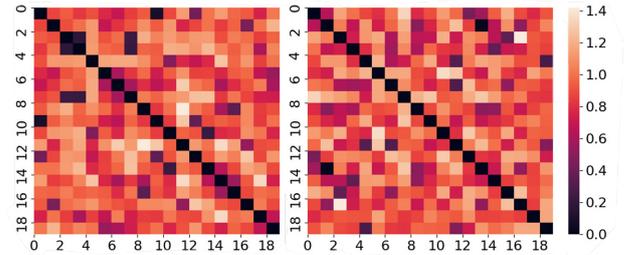

Fig. 4. *Averaged distance matrices of 19 test musical pieces calculated with the learned music similarity metrics focusing on bass (left) and guitar(right). These distance matrices illustrate the distance between the centroids of each musical piece. For example, a value in column 0 and row 10 represents the distance between the centroids of music no. 0 and music no. 10. The distance matrices are averaged over five trials. The darker the color between two musical pieces, the more similar the two pieces are.*

TABLE III
*Correlation results on the averaged distances over 19 test musical pieces between different music similarity metrics.*

(a) Correlation coefficients

| instrument | mix | drums | bass | piano | guitar |
|---|---|---|---|---|---|
| mix | 1 | 0.23 | 0.11 | 0.12 | 0.32 |
| drums | | 1 | -0.0016 | 0.16 | 0.025 |
| bass | | | 1 | 0.021 | 0.021 |
| piano | | | | 1 | 0.12 |
| guitar | | | | | 1 |

(b) Spearman's rank correlation coefficients

| instrument | mix | drums | bass | piano | guitar |
|---|---|---|---|---|---|
| mix | 1 | -0.020 | 0.088 | -0.082 | 0.085 |
| drums | | 1 | 0.076 | 0.018 | 0.086 |
| bass | | | 1 | 0.038 | 0.091 |
| piano | | | | 1 | 0.054 |
| guitar | | | | | 1 |

128-dimensional feature representations to two-dimensional representations. We can see that feature representations from the same musical piece were concentrated, forming clusters.

Figure 4 shows an example of the average distance matrices. We can visually confirm the difference between feature representation spaces learned with different instrumental sounds from these examples. Table III shows the correlation coefficients and the Spearman's rank correlation coefficients. Both results with values very close to zero indicated that unique similarity metrics were learned by using individual instrumental sounds.

*2) Subjective Evaluation:* Figure 5 shows the results of the subjective experiment. "True" and "False" are used to denote cases of the positive and negative musical pieces being selected, respectively. "+" and "-" denote that the listener was confident and not confident in the choice, respectively. Including less confident choices, listeners selected more than 50% of the positive samples in all the cases, where the percentages in the mixed, bass, and guitar cases were higher than 70% as shown in Table IV. These results indicated that the music similarity metrics learned with the proposed method focusing on a specific type of sound could find more perceptually similar segments in terms of the focused perspective, and they well corresponded to the perception of human senses.

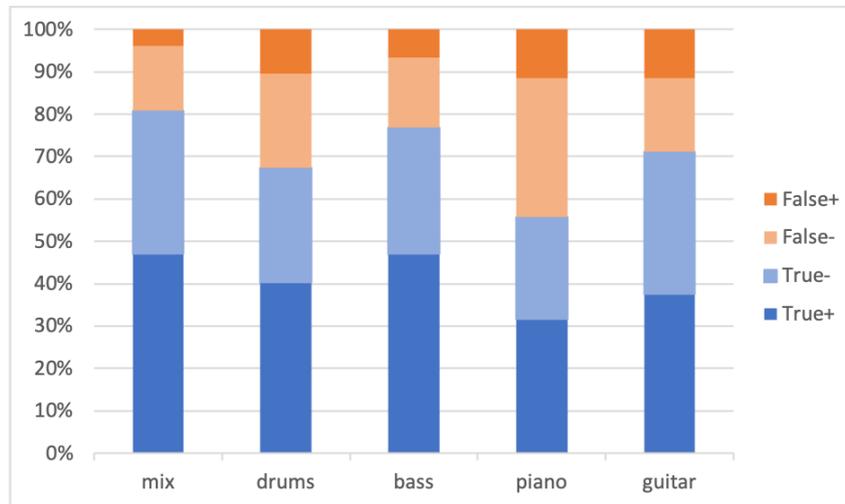

Fig. 5. *Result of subjective evaluation on correspondence of music similarity metrics to perceived similarity.*

TABLE IV
*True rates and 95% confidence intervals as a result of subjective evaluation*

| instrument | mix | drums | bass | piano | guitar |
|---|---|---|---|---|---|
| true rate[%] | 80.8±7.6 | 67.3±9.1 | 76.9±8.1 | 55.8±9.6 | 71.2±8.7 |

## V. CONCLUSION

In this paper, we proposed a music similarity calculation method focusing on individual instrumental sound sources. The proposed method learns similarity metrics using deep metric learning with a triplet loss. Experiments showed that it was possible to learn the similarity metrics focusing on different musical instruments. In addition, we found that similarity metrics learned using some instrumental sounds led to more accurate results than that learned using the entire piece of music. However, the performance degraded when learning with the separated instrumental sounds. The subjective experimental results revealed that the proposed music similarity metrics well corresponded to perceptual similarity. Future work includes application to songs that include vocals, and joint optimization of metric learning and music source separation.

**Acknowledgement:** This work was partly supported by JST CREST Grant Number JPMJCR19A3, Japan.